# Boundary-Induced Biases in Climate Networks of Extreme Precipitation and Temperature

Behzad Ghanbarian[1,2,3*], Victor Oladoja[4], Kehinde Bosikun[5], Tayeb Jamali[6], Jürgen Kurths[7,8]

[1] iResearchE[3] Lab, Department of Earth and Environmental Sciences, University of Texas at Arlington, Arlington 76019 TX, USA

[2] Department of Civil Engineering, University of Texas at Arlington, Arlington TX 76019, United States

[3] Division of Data Science, College of Science, University of Texas at Arlington, Arlington TX 76019, United States

[4] Department of Earth Sciences, University of Connecticut, Storrs CT 06269 USA

[5] Department of Chemical, Biochemical and Environmental Engineering, University of Maryland at Baltimore County, Baltimore MD 21250 USA

[6] Broad Institute of MIT and Harvard, Cambridge, MA, USA

[7] Potsdam Institute for Climate Impact Research, D-14412 Potsdam, Germany

[8] Department of Physics, Humboldt University of Berlin, D-12489 Berlin, Germany

[*] Corresponding author's Email address: ghanbarianb@uta.edu

**Abstract**

Climate networks have recently emerged as a powerful tool for the spatiotemporal analysis of climatic and atmospheric phenomena. When constructed over domains smaller than the global scale, such as the contiguous United States (CONUS), these networks are susceptible to spatial boundary effects, which can bias network metrics, particularly near the domain edges. To address this issue, two surrogate-based correction methods, (1) subtraction and (2) division, have been widely applied in the literature. In the subtraction method, an original network measure is adjusted by subtracting the expected value obtained from a surrogate ensemble, whereas in the division method, it is normalized by dividing by this expected value. However, to the best of our knowledge, no prior study has assessed whether these two correction approaches yield statistically different results. In this study, we constructed complex networks of extreme precipitation and temperature events (EPEs and ETEs) across the CONUS for both summer (June-August, JJA) and winter (December-February, DJF) seasons. We computed key network metrics degree centrality (DC), clustering coefficient (CC), mean geographic distance (MGD), and betweenness centrality (BC) and applied both correction methods. Although the corrected spatial patterns generally appeared visually similar, statistical analyses revealed that the network measures derived from the subtraction and division methods were significantly different at the 95% confidence level. Across the CONUS, network hubs of EPEs were primarily concentrated in the northwestern United States during summer and shifted toward the east during winter, reflecting seasonal differences in the dominant atmospheric drivers. In contrast, the ETE networks showed strong spatial coherence and pronounced regional teleconnections in both seasons, with higher connectivity and longer synchronization distances in winter, consistent with large-scale circulation patterns such as the Pacific–North American and North Atlantic Oscillation modes. Our results indicated that the network metrics CC and MGD were more sensitive to the correction methods than the DC and BC, particularly in the EPE networks.

# 1. Introduction

The Earth's climate system exhibits a highly complex structure, characterized by numerous components interacting through nonlinear processes (Rial et al., 2004). To analyze such complexity, a wide range of methodologies has been developed, including an array of statistical techniques designed to capture dependencies, patterns, and variability across spatial and temporal scales. Notable examples are geostatistics (Goovaerts, 2000), wavelet coherence (Labat, 2005), empirical orthogonal function analysis (Pritchard & Somerville, 2009) and principal component analysis (Tadić et al., 2019) that were widely applied to identify dominant spatial and/or temporal patterns. For instance, Svensson (1999) analyzed rainfall data (1967-1986) from East Central China located in the east Asian monsoon region using the empirical orthogonal function analysis. His results showed that an elongated spatial precipitation pattern was responsible for most of the variance in the studied area. He also found that the pattern direction varied from west-to-east to southwest-northeast.

Recurrence quantification analysis, originally proposed by Zbilut & Webber Jr (1992), was also proposed to study recurring states in climate systems and identify regime shifts and transitions (Riedl et al., 2015; Trauth, M. H. et al., 2019). Unlike power spectral and wavelet analyses, the recurrence quantification analysis is not restricted to periodic variations (Boers et al., 2021). Banerjee et al. (2020) studied mean daily streamflow values of the Mississippi River from the Clinton station in Iowa and then computed flood event series by determining extreme values from the streamflow series (events above the $99^{th}$ percentile threshold). To analyze the recurrence of extreme flood events, they proposed a distance measure, based on a modified edit distance method, included the concept of time delay and found deterministic patterns in the occurrence of extreme flood events.

In recent years, concepts from complex network theory and network science have been applied to atmospheric and climate sciences (Donges et al., 2009; Ludescher et al., 2021; Meng et al., 2018; Yamasaki et al., 2008; Zou et al., 2018). Various hydrological and climatic variables, such as temperature, precipitation, pressure, or streamflow, can serve as the basis to construct climate networks. Network-based methods enable the investigation of spatial interrelationships across regions and scales, ranging from local (Agarwal et al., 2018) to global (Boers et al., 2019). They also facilitate the exploration of how energy, moisture, or information is transferred across the climate system, offering a powerful framework for understanding complex geophysical dynamics (Konapala & Mishra, 2017; Zou et al., 2018).

Applications of climate networks have provided us valuable insights for analyzing, modeling, understanding, and even predicting a wide range of climatic phenomena (Fan et al., 2021). For instance, in a recent study, Ludescher et al. (2021) demonstrated the practical applications and predictive power of complex network theory in climate science. By leveraging network-based indicators derived from climate data, they successfully forecasted a range of high-impact events, including El Niño episodes, droughts in the central Amazon, extreme rainfall in the eastern Central Andes, and variations in the Indian summer monsoon. Notably, their analysis showed that complex network measures could significantly enhance predictive skill. For example, six of the seven most severe droughts in central Amazon over the past four decades were accurately predicted using their network-based approach, highlighting capabilities in anticipating critical climate extremes.

Within the framework of climate networks, geographic locations represent nodes and the level of similarity or statistical dependency between time series at different locations represents links. In spatially confined embedded networks, nodes near the boundary are likely to form connections with places outside the domain of study. This is especially true for complex networks of climatic variables, such as extreme temperature and precipitation events. For example, Oladoja et al. (2025) analyzed extreme precipitation events across North America by means of complex network theory. They found that super nodes (or hubs) in Montana, Wyoming and Idaho in the United States were linked to those across the border in Alberta and Saskatchewan in Canada during the summer season (see their Fig. 3a).

The spatial boundary effect and its correction have been a challenge in geospatial analyses (Griffith, 1985). In climate networks constructed at sub-global scales, such as those over the contiguous United States (CONUS), the geographic location of a node, particularly if it is located near spatial boundaries, can affect the accuracy of network metrics and interpretations of climate connectivity patterns. Nodes situated near domain edges may appear structurally distinct not because of actual physical disconnection, but due to the limited number of neighboring nodes beyond the boundaries. To mitigate these boundary-induced biases, two surrogated-based correction methods (subtraction and division) have been widely used in the climate network literature. These methods aim to adjust network metrics by accounting for reduced connectivity near edges and by estimating the network structure expected purely from spatial embedding. Both subtraction and division approaches involve generating ensembles of spatially embedded surrogate networks that preserve the geographic positions of nodes and the overall degree distribution, while randomizing the connectivity structure. The subtraction method, introduced by Rheinwalt et al. (2012), corrects a given network metric by subtracting the surrogate ensemble mean from the original value. Alternatively, the division method, proposed by Boers et al. (2013), normalizes each network metric by dividing the original value by the surrogate ensemble mean.

Although both correction methods have been independently employed to reduce spatial boundary effects in climate networks, a systematic comparison of their performance has not yet been conducted, to the best of the authors' knowledge. Therefore, the main objectives of this study are to: (1) construct climate networks of extreme precipitation and temperature events (EPEs and ETEs) across the CONUS for both summer and winter seasons; (2) apply surrogate-based subtraction and division methods to correct for spatial boundary effects; and (3) critically compare the results of the two correction techniques.

**2. Precipitation and temperature data**

The precipitation and temperature data used in this study were collected from the Climate Prediction Center (CPC) database, provided by NOAA/OAR/ESRL PSD (USA) and publicly available at https://psl.noaa.gov. This gridded database contains daily precipitation as well as minimum and maximum temperature records since 1979. However, in this study, we used the precipitation temperature data from 1991 to 2020 (thirty years). The spatial resolution in the CPC database is 0.5° × 0.5°, which resulted in 3,276 grid points within the contiguous United States (CONUS). We separately extracted records for all summer (June, July, and August; hereafter JJA) and winter (December, January, and February; hereafter DJF) seasons over the three-decade period.

For the precipitation data, we calculated the EPEs as daily values exceeded the 95$^{th}$ percentile of the local precipitation distribution at each grid point and determined event series by assigning a value of one to the extreme events and zero to all others (Boers et al., 2019). We should point out that Boers et al. (2019) and Jamali et al. (2023) determined the 95$^{th}$ percentile threshold based on wet days defined as events with more than 1 mm of precipitation. In this study, however, we included all precipitation events above zero to calculate the 95$^{th}$ percentile threshold at each node.

For the temperature data, we analyzed the maximum temperature records and defined ETEs as days with values exceeded the 95$^{th}$ percentile threshold at each grid point in summers. However, minimum temperature data were used, and ETEs were identified as days with values that fell below the 5$^{th}$ percentile at each grid point in winters.

Using the corresponding threshold at each grid point, we converted the temperature and/or precipitation time series into event series, where a value of one indicated the occurrence of an extreme event and zero denoted its absence. We should point out that for the EPEs and ETEs, the 95$^{th}$ and/or 5$^{th}$ percentile thresholds were computed individually at each grid point to account for local climate variability. In the case of consecutive extreme days, only the first event in each sequence was retained to prevent temporal clustering bias, consistent with the approach of Boers et al. (2019).

**3. Climate networks**

Climate networks represent a distinct class of complex systems embedded on the Earth's surface. They consist of nodes, corresponding to grid points, connected by links. In this study, we constructed undirected and unweighted networks of EPE and ETE separately for the summer and winter seasons using the EPE and ETE series. In what follows, we explain how links were determined and how the EPEs and ETEs networks were constructed in particular within the geographic boundaries of the CONUS.

**3.1. Event synchronization method**

We employed a modified version of the event synchronization (ES) method, developed by Rheinwalt et al. (2016), to determine the level of similarity between any two nodes across the CONUS. In the original ES method, Quiroga et al. (2002) proposed normalizing the ES measure to account for differences in event rates. Rheinwalt et al. (2016), however, demonstrated that this normalization does not fully eliminate the influence of event rate variability and proposed to apply a uniform significance threshold to all pairs, as an alternative.

Let $\{t_m^i\}_{m=1}^{N_i}$ be the event series at node $i$ with length $N_i$ and $\{t_n^j\}_{n=1}^{N_j}$ at node $j$ with length $N_j$. Then, one can calculate the $ES^{i,j}$, degree of associable events between nodes $i$ and $j$, by counting the number of temporal delays $\Delta_{m,n}^{i,j}$ that satisfies $|\Delta_{m,n}^{i,j}| < \tau_{m,n}^{i,j}$ (dynamic local time scale depending on the events) given $|\Delta_{m,n}^{i,j}| \leq \tau_{max}$ (a lag time; the maximum delay allowed between two extreme events from two different nodes). The $\Delta_{m,n}^{i,j}$ and $\tau_{m,n}^{i,j}$ are given by:

$$\Delta_{m,n}^{i,j} = t_n^j - t_m^i \tag{1}$$

$$\tau_{m,n}^{i,j} = \frac{1}{2} \times min\{\Delta_{m,m-1}^{i,i}, \Delta_{m,m+1}^{i,i}, \Delta_{n,n-1}^{j,j}, \Delta_{n,n+1}^{j,j}\} \tag{2}$$

The temporal delay $\Delta_{m,n}^{i,j}$ is restricted to a maximum threshold $\tau_{max}$ (Boers et al., 2013) to avoid unrealistic synchronizations. In this study, we set $\tau_{max} = 0$ for both precipitation and temperature. This means that extreme events were considered synchronized only if they occurred on the same day.

To account for potential biases arising from variations in the number of events at different nodes, an issue highlighted by Boers et al. (2016), we applied a null model-based approach, which assumes that event correlations between any two nodes are purely random (Boers et al., 2019). This method enables an objective assessment of whether the observed synchronization between two nodes is statistically significant. For a given pair of nodes $i$ and $j$, we computed the event synchronization value, $ES^{i,j}$, based on their event time series. To assess its statistical significance, we generated a null distribution by randomly shuffling the event series $t^i_{1,\ldots,m}$ and $t^j_{1,\ldots,n}$ independently 1,000 times. In each iteration, a new $ES$ value was calculated and added to the null model distribution. We then used the 99.5th percentile of the null distribution as the significance threshold. A statistically significant link was established between nodes $i$ and $j$ if the observed $ES^{i,j}$ was greater than or equal to this threshold. This means that the probability of a purely random link being accepted was at most 0.005.

**3.2. Climate network metrics**

To effectively characterize spatial patterns in the constructed climate networks, we used a range of network metrics. These measures provide insights into the structural and spatial organization of the networks. In this section, we briefly introduce four key network metrics that have been widely used in the analysis of complex systems, particularly climate networks, to quantify node-level and global properties.

**Degree Centrality (DC ≥ 0)**: to quantify the local connectivity of a node, one may compute the degree centrality by counting the number of links that it shares with other nodes as follows

$$DC_i = \sum_{j=1}^{n} A_{(i,j)} \tag{3}$$

where $A_{i,j}$ is an element of the adjacency matrix, equal to 1 if there is a link between nodes $i$ and $j$, and 0 otherwise, and $n$ is the total number of nodes in the network. A higher degree value indicates that the node is more centrally located within the network of synchronized extreme events.

**Clustering Coefficient (0 ≤ CC ≤ 1):** to measure the extent to which the neighbors of a given node are themselves interconnected, one may calculate the clustering coefficient, reflecting the likelihood that two nodes connected to a common node are also directly connected to each other. The $CC$ captures the local cohesiveness or link density around each node and given by

$$CC_i = \frac{2L_i}{DC_i(DC_i - 1)} \tag{4}$$

where $L_i$ is the number of links that exist between the $DC_i$ neighbors.

**Mean Geographic Distance (MGD ≥ 0)**: a node-based metric that quantifies the average spatial distance between a given node and all its connected neighbors. MGD provides insight into whether a node tends to synchronize with nearby or distant regions in terms of extreme precipitation or temperature patterns and defined as

$$MGD_i = \frac{1}{DC_i} \sum_{j=1}^{n} D_{(i,j)} A_{(i,j)} \tag{5}$$

where $D_{(i,j)}$ is the geographic distance between nodes $i$ and $j$. A lower MGD value indicates that a node is primarily linked to nearby locations, while a higher value suggests stronger long-range spatial associations.

**Betweenness centrality (BC ≥ 0):** this is a node-based metric widely used to identify dynamic and structural importance within climate networks. BC highlights a node's role as a bridge or intermediary within the network, unlike DC, which emphasizes the number of direct connections that a node has. BC is given by

$$BC(v_i) = \frac{2 \sum_{v_s \neq v_i \neq v_t} \frac{\sigma_{st}(v_i)}{\sigma_{st}}}{(n-1)(n-2)} \quad (6)$$

where $\sigma_{st}(v_i)$ is the number of $\sigma_{st}$ that passes through the node $v_i$ and $\sigma_{st}$ is the number of shortest paths between nodes $v_s$ and $v_t$. BC quantifies how often a node lies on the shortest paths between pairs of other nodes, assessing its potential influence on the flow or propagation of information across a network.

## 4. Spatial boundary effect and its correction

In climate networks constructed at scales smaller than global (e.g., CONUS at the national scale), the geographic location of a node, particularly its proximity to spatial boundaries of the study area, can influence the accuracy and interpretation of network metrics. Nodes near boundaries have fewer potential neighbors within the study domain compared to interior nodes. This creates an artificial constraint on connectivity that does not reflect the true climate system. To mitigate such boundary effects, correction methods have been proposed in the literature. In what follows, we describe two spatial boundary correction methods widely used in climate networks.

### 4.1. Surrogate-based subtraction method

To correct climate networks for spatial boundary effects, Rheinwalt et al. (2012) subtracted the expected value of the measure computed from an ensemble of surrogate networks from the original network measure. This correction is applied at each individual node in the network They generated surrogate networks by randomizing the placement of links while preserving the spatial link probability $p(\Delta)$ i.e., the probability that two nodes separated by the distance $\Delta$ are connected. Rheinwalt et al. (2012) qualitatively evaluated their approach using one artificial random network. Following Rheinwalt et al. (2012), Rheinwalt (2015), Gao et al. (2023) and others corrected their constructed complex networks using the subtraction method.

In this study, we generated an ensemble of 1000 surrogate networks and, at each grid point, calculated the arithmetic average of each network measure across the ensemble, denoted as $\langle M_i^{sur} \rangle$. To correct for boundary effects, we then adjusted the original measure at the grid point $i$ by subtracting the surrogate average, resulting in the corrected measure: $M_i^{cor} = M_i - \langle M_i^{sur} \rangle$. Subsequently, we rescaled the corrected measure using the min-max normalization across all grid points to map values within the interval [0,1]: $M_i^{norm} = (M_i^{cor} - M_{min}^{cor})/(M_{max}^{cor} - M_{min}^{cor})$ where $M_{min}^{cor}$ and $M_{max}^{cor}$ are the minimum and maximum values of $M_i^{cor}$ across all grid points. This correction ensures that spatial artifacts due to edge proximity are minimized, and the resulting network measures reflect the intrinsic topological properties of the system more accurately.

**4.2. Surrogate-based division method**

Another widely-applied spatial boundary correction technique in the climate network literature is the division method, first proposed by Boers et al. (2013). Notable studies that applied this approach include Boers (2015), Gupta et al. (2021), and Oladoja et al. (2025), among others. This approach is conceptually similar to the subtraction method; both aim to reduce biases introduced by edge effects using surrogate networks. However, the division method involves normalizing the original network measure at each node by its expected value derived from an ensemble of surrogate networks ($M_i^{cor} = \frac{M_i}{\langle M_i^{sur} \rangle}$).

To correct the constructed EPE and ETE networks via the division method, we used the generated 1000 surrogated networks, computed the arithmetic average for each network measure from the surrogates and calculated the value expected from the spatial embedding alone. Next, we divided the original network metric by the expected value from the surrogate ensembles.

**5. Statistical analysis**

After correcting a network measure as outlined in the preceding section, two corrected sets of node-specific values were obtained: one generated through subtraction and the other through division. Both sets of corrected values were subsequently analyzed using the aforementioned statistical tests.

To evaluate whether the surrogate-based subtraction and division methods differ significantly in correcting climate network metrics for spatial boundary effects, we conducted two complementary statistical tests: (1) the paired t-test and (2) the Kolmogorov–Smirnov (K-S) test. The paired t-test examined differences in central tendency, with the null hypothesis that the mean values of network metrics corrected using the subtraction and division methods are equal. The K-S test assessed distributional differences, with the null hypothesis that the corrected values from both methods originate from the same underlying probability distribution. A significance level of 0.05 (corresponding to a 95% confidence level) was adopted for both tests.

## 6. Results and Discussion

In this section, we first present and discuss the results for the EPE networks, followed by those for the ETE networks.

### 6.1. EPE climate networks

The DC of a grid point, calculated using Eq. (3), represents the number of other grid points to which it is connected through synchronized EPEs. The DC results are presented in Fig. 1 for both the original and boundary-corrected EPE networks during the summer (left column; JJA) and winter (right column; DJF) seasons. Figure 1a displays the spatial distribution of DC in the uncorrected EPE network for the summer season. The networks corrected using the surrogate-based subtraction and division methods are shown in Figs. 1b and 1c, respectively. While the spatial patterns resulting from the two correction methods appear qualitatively similar, their outcomes are indeed statistically different. A paired t-test revealed that the mean DC value corrected using the subtraction method differs significantly from that obtained using the division method (Table 1). The K-S test further confirmed that the DC distributions resulted from the subtraction and division methods were significantly different for the summer season, with a p-value less than 0.05 (see Table 1).

Regardless of the correction method applied, we identified network hubs in the northwestern region of the CONUS, primarily across Montana, Idaho, and Wyoming (Figs. 1b and 1c). These findings are generally consistent with those reported by Jamali et al. (2023), who investigated the spatiotemporal patterns of extreme precipitation across the CONUS and found hub nodes mainly concentrated in Montana (see their Fig. 4a). However, our results indicate a broader spatial extent of hub regions in the northwestern

United States. This discrepancy may be attributed to a key methodological difference: although both studies used the 95th percentile threshold to define extreme precipitation events, the definition of wet days varied. In our study, wet days were defined as those with precipitation greater than 0 mm, whereas Jamali et al. (2023) considered only days with precipitation exceeding 1 mm. This difference in thresholding likely affected the identification of extreme precipitation events and, consequently, the resulting network structure.

In both corrected networks (Figs. 1b and 1c), relatively high DC values were also observed along the Appalachian Mountains in the eastern CONUS. This pattern may be attributed to enhanced latent heat flux in the region, which promotes deep convection, facilitates moisture condensation, and ultimately increases the likelihood of synchronized EPEs (Mondal et al., 2020). Furthermore, Li et al. (2021) reported that the local topography and land-sea heating contrast have a strong influence on the summer precipitation patterns in the Mid-Atlantic region through mountain-valley breezes that blow from the Appalachian Mountains to the nearby low-elevation areas and land-sea breezes that blow from the ocean towards land.

Fig. 1d presents the uncorrected DC across the CONUS during the winter season (DJF). The spatial boundary-corrected DC values, obtained using the surrogate-based subtraction and division methods, are shown in Figs. 1e and 1f, respectively. In contrast to the results of the summer season (Figs. 1b and 1c), the hubs during the winter season are concentrated in the east CONUS. As can be seen, Figs. 1e and 1f exhibit similar spatial patterns, with higher DC values generally concentrated in the eastern CONUS and lower values in the western region. However, the magnitudes of DC differ between the two correction methods, with Fig. 1f (division method) yielding systematically lower values than Fig. 1e (subtraction method). This discrepancy is likely due to differences in how the division method normalizes the network structure. We found that when the network links are randomly shuffled to generate a surrogate ensemble, a fraction of nodes may become isolated (degree zero). In an ensemble of surrogate networks, this means that some nodes could have an average degree much smaller than one. In such cases, in the division-based correction, these small denominators lead to disproportionately large corrected values. Consequently, when the corrected values are mapped onto the [0,1] interval, the presence of extremely large maxima compresses the majority of values toward zero, leaving only a small fraction of large values near one. By contrast, the subtraction-based correction addresses this issue more smoothly. This behavior explains why a statistically significant difference is observed between the DC values obtained from the subtraction and division methods.

Our statistical analyses further support the distinction between the two correction methods. Both the paired t-test and the K-S test yielded small p-values ($\ll 0.05$), as reported in Table 1, indicating statistically significant differences between the DC values derived from the subtraction and division methods. Specifically, the t-test revealed a significant difference in mean DC values, while the K-S test showed that the DC values from the two methods follow different distributions.

Figure 2 presents the spatial distribution of the CC across the CONUS for both the summer and winter seasons. For the summer season, the uncorrected CC values are shown in Fig. 2a, while Figs. 2b and 2c display the corrected CC values obtained using the surrogate-based subtraction and division methods, respectively. Although the overall spatial patterns produced by both correction methods appear qualitatively similar and consistent, the statistical analysis reveals significant differences between them. As shown in Table 1, a paired t-test indicates that the mean CC values obtained from the subtraction method differ significantly from those derived using the division method (p-value ≪ 0.05). Furthermore, the results from the KS test demonstrate that the two sets of corrected CC values do not originate from the same distribution (p-value ≪ 0.05), further confirming the statistical distinction between the two correction approaches.

Generally speaking. Fig. 2c displays regions with relatively higher CC values compared to Fig. 2b. Figs. 2b and 2c both show very high CC values in California. Both approaches revealed relatively high CC in regions, such as south Texas, northern California and US northwest and northeast. Our results are generally consistent with those of Mondal et al. (2020) who investigated spatiotemporal dynamics of EPEs in the United States during summers by analyzing precipitation data from 1979 to 2017.

The uncorrected and corrected CC values within the CONUS for the winter season are presented in Figs. 2d-f. Overall, compared to the summer season (Figs. 2a-2c), the EPEs during winters exhibited greater spatial coherence (Figs. 2d-2f). Figs. 2e and 2f show very high CC values in eastern Colorado and western Arizona. Similar results were reported by Oladoja et al. (2025) who analyzed EPEs in North America (see their Fig. 6c).

Similar to the summer season results (Figs. 2b and 2c), both the surrogate-based subtraction and division methods produced spatially consistent corrections, effectively addressing the boundary effect. Despite these visual similarities, statistical analyses revealed significant differences between the two correction approaches. Specifically, the paired t-test and K-S test results (Table 1) yielded p-values ≪ 0.05, indicating that the mean CC values from the two methods are significantly different and that their underlying distributions are not statistically equivalent. These findings confirm that, although the subtraction and division methods yield comparable spatial patterns, they differ in the magnitude and distribution of the corrected CC values.

The spatial patterns of the synchronization distance of EPEs across the CONUS can be evaluated using the MGD. Regions with low MGD values indicate areas characterized by short-range synchronizations, whereas high MGD values suggest the presence of long-range synchronizations or teleconnections. Figure 3 presents the spatial distribution of the mean geographic distance (MGD) for both summer and winter seasons, including the uncorrected and corrected versions. It is important to note that, following the application of spatial boundary correction methods, the MGD values become dimensionless due to normalization with surrogate-based expectations. In the maps, brighter regions correspond to areas dominated by short-range synchronizations, whereas darker regions indicate the presence of long-range synchronizations, often associated with teleconnections.

As shown in Figs. 3b and 3c, both correction methods yielded generally low to moderately low MGD values (< 0.4) across the CONUS during the summer season. This suggests that most atmospheric processes responsible for extreme precipitation events (EPEs) in summer are localized and do not exhibit large-scale propagation across the CONUS. Notable exceptions include certain regions in Florida, Texas, and California, where relatively higher MGD values were observed, patterns consistent with the findings of Jamali et al. (2023). Although the spatial patterns produced by the subtraction and division methods appear similar, statistical analyses revealed significant differences between the two. Specifically, the paired t-test indicated that the mean MGD values obtained from the surrogate-based subtraction method were significantly different from those produced by the division method (p-value ≪ 0.05; Table 1). Furthermore, the K-S test showed that the MGD values derived from the two methods do not follow the same distribution (p-value ≪ 0.05; Table 1).

The MGD results for the winter season are presented in Figs. 3d-3f. Compared to the summer season (JJA), grid points with high MGD values are more widely distributed across the CONUS during the winter season (DJF). This is likely due to the influence of larger-scale synoptic systems that dominate precipitation generation during winter, resulting in more spatially extensive and coherent extreme events (Jamali et al., 2023).

Both correction methods (Figs. 3e and 3f) show very high MGD values in some regions in New England, Southeast, Colorado, New Mexico and northern California in the CONUS. Generally speaking, east CONUS has higher MGD values compared to west CONUS. During winters, teleconnections in the eastern CONUS are mainly due to large-scale atmospheric circulation patterns (e.g., North Atlantic Oscillation, El Niño–Southern Oscillation and Pacific North America pattern) and moisture transport mechanisms (e.g., arm, moist air masses from the Gulf of Mexico and the western Atlantic) that link distant climate anomalies.

BC quantifies the importance of a node in facilitating the transfer of EPEs among different regions within the complex EPE network. It measures the extent to which a node serves as an intermediary in the long-range spatial propagation of EPEs and, thus, highlights nodes that play a pivotal role in connecting otherwise distant areas. Fig. 4 presents the spatial distribution of uncorrected and corrected BC values across the CONUS for the summer (JJA) and winter (DJF) seasons. The BC values typically span a wide range and generally are small in magnitude (Figs. 4a and 4d). We, therefore, applied a logarithmic transformation, $log\ (BC + 1)$, to enhance visual interpretation.

The BC results corrected via the surrogate-based subtraction and division methods are shown in Figs. 4b and 4c, respectively, for the summer season. Differences between the two approaches are obvious. The surrogate-based subtraction method revealed few locations with high BCs (> 0.7) (Fig. 4b). We found high BCs concentrated in a region covering northern Colorado, Wyoming, and western Montana. Other areas with relatively high BCs include eastern Nevada, southern and northern Utah, southcentral Texas, northeast Georgia, and northeastern Kentucky. Our results align with those of Oladoja et al. (2025), who applied complex network theory to analyze EPEs across North America (see their Fig. 7a), and Mondal et al. (2020), who focused on summer EPEs in the United States (see their Fig. 2a). Using the surrogate-based division method, however, we found a broader spatial distribution of high BC values across the CONUS (Fig. 4b vs. Fig. 4c). Compared to the subtraction method, the division approach revealed additional regions, such as the East and West Coast, southwest Texas, Florida, etc., exhibiting moderate to high BC values (Fig. 4c). Unsurprisingly, the paired t-test and K-S test showed statistically significant differences between the two correction methods. The paired t-test (p-value << 0.05) showed that the mean BC value derived from the subtraction method is significantly different from that obtained using the division method. Similarly, the K-S test (p-value << 0.05) revealed that the BC distributions corresponding to the two correction methods originate from statistically distinct distributions.

For DJF, the results from the surrogate-based subtraction method are presented in Fig. 4e, while those from the division method are presented in Fig. 4f. Similar to the summer season, the division method yields a more widespread spatial distribution of moderate to high BC values across the CONUS compared to the subtraction method (Fig. 4e vs. Fig. 4f). We found high BCs (>0.7) in a region patch extending from northern New Mexico to northern Colorado (Fig. 4e) using the subtraction method. The surrogate-based division method revealed high BC values in western California, western Oregon, southern Arizona, eastern Texas, Louisiana, southern Florida, eastern Oklahoma, eastern Virginia, northern Wisconsin, northern New York, and northwestern Maine, indicating these regions may serve as critical connectors within the winter EPEs network. Our findings are consistent with Banerjee et al. (2023), who analyzed extreme winter precipitation in the United States using complex network theory and the edit distance method, based on ERA5 reanalysis precipitation data (see their Fig. 4b). Both the t-test (p-value << 0.05) and K-S test (p-value << 0.05) showed that the mean BC values derived from the two methods are significantly different, and that the BC distributions obtained from the subtraction and division methods originate from statistically distinct distributions.

**6.2. ETE climate networks**

Fig. 5 presents the DC values for the grid points within the CONUS during the summer and winter seasons, before and after correcting for the spatial boundary effect. Although there exist similarities between Figs. 5b and 5c, differences between the two correction approaches are distinct for the summer season. For instance, both methods revealed very high DC values in New England and the Southwest in the CONUS. The complex topography of the Appalachian Mountains and the Cascade Range plays a significant role in shaping local climate (temperature) conditions in eastern CONUS and the Pacific Northwest, respectively. Mountains are frequently regarded as climate change hotspots, as they often experience effects of climate change more rapidly and intensely than lower-elevation regions. As such, they serve as early indicators of broader climatic shifts that may eventually manifest across larger spatial scales (Pepin et al., 2022).

Results of the paired t-test and K-S test for the DC during the summer season are summarized in Table 1. As shown, both tests yielded p-values much smaller than 0.05, indicating statistically significant differences between the two correction methods. Specifically, the paired t-test revealed that the mean DC values obtained using the subtraction method differ significantly from those derived using the division method. Similarly, the K-S test indicated that the distributions of DC values resulting from the two correction approaches are statistically distinct, further confirming the methodological divergence in their outcomes.

Figure 5d displays the spatial distribution of uncorrected DC values across the CONUS during the winter season. These values range from 0 to approximately 2700, representing a significantly broader range than those observed in the summer season ($0 < DC < 1050$; see Fig. 5a). The spatial distributions of the corrected DC values obtained using the surrogate-based subtraction and division methods are shown in Figs. 5e and 5f, respectively. The southern Great Plains have moderately high DC during the winter season, a pattern more pronounced in the subtraction method (Fig. 5e). The region's flat topography and lack of major water bodies make it susceptible to the influence of different air masses, especially those from the Northwest (Rosenberg, 1987). Furthermore, southward advancement of Arctic fronts often results in temperature drops and severe cold events (Singh et al., 2016). These temperature extremes are further intensified by the jet stream fluctuation, which can allow cold air masses penetrate into the region, resulting in significant temperature extremes (Ma & Chang, 2017). Both correction approaches highlight regions with elevated DC values in New England, the Great Lakes, the South, and the western part of the Northwest, indicating higher synchronization of ETEs across these regions in winters. The northeastern CONUS is frequently influenced by cyclonic systems (e.g., nor'easters), which are associated with substantial temperature fluctuations and extreme winter weather conditions (Low et al., 2022). In the northwestern CONUS, temperature variability is strongly modulated by the Pacific-North American (PNA) pattern. Certain phases of the PNA can cause shifts in the polar jet stream, facilitating cold air outbreaks and leading to ETEs (Ning & Bradley, 2015).

Statistical analyses using both the paired t-test and the K-S test revealed significant differences between the two correction methods. Specifically, the p-values reported in Table 1 were much smaller than 0.05, indicating strong statistical significance. The paired t-test confirmed that the mean DC value determined using the subtraction method was significantly different from that calculated using the division method. Furthermore, the K-S test results demonstrated that the DC distributions resulted from the two correction approaches originated from statistically distinct distributions.

The CC quantifies the level of interconnectedness among a node's neighbors within the ETE complex network. Nodes with high CC values tend to be well-connected to their immediate neighbors, reflecting a high degree of spatial coherence in the occurrence of ETEs. Figure 6 presents the CC results including both uncorrected and spatial-boundary corrected values for the summer and winter seasons. Notably, Figs. 6b and 6c reveal very high CC values in west Texas and eastern New Mexico, south of the Rocky Mountains. As will be shown later, this region exhibits very low MGD values, suggesting that these high

CC values are associated with highly localized summer ETEs and heatwaves (Bosikun et al., 2025). This region also exhibits very low DC, as shown in Fig. 5, indicating limited synchronization of ETEs with other areas. Interestingly, west Texas has been previously identified as a vulnerable region prone to intense and persistent drought conditions (Rajsekhar et al., 2015). We should also point out that our CC results are in general agreement with those of Mondal & Mishra (2021) who analyzed the network structure and investigated the propagation characteristics of summer heatwaves across the CONUS using complex network theory and daily maximum temperature data (see their Fig. 2c).

Although the spatial distributions of CC during summers are visually similar between the two correction methods (i.e., Fig. 6b vs. Fig. 6c), our statistical analyses revealed strong and significant differences. Specifically, the paired t-test showed that the mean CC values derived from the surrogate-based subtraction and division methods differ significantly (p-value << 0.05; Table 1). Similarly, the K-S test indicated that the distributions of CC values from the two correction methods originate from statistically distinct populations (p << 0.05; Table 1).

Figs. 6d-6f present the CC results for the winter season. The surrogate-based subtraction and division methods produced again visually similar spatial patterns, similarly correcting for the spatial boundary effect (Figs. 6e and 6f). This observation is supported by the statistical analyses. Specifically, the paired t-test yielded a p-value of 0.995 (Table 1), indicating no statistically significant difference between the mean CC values obtained from the two correction methods at the 0.05 significance level. However, the K-S test returned a p-value of 0.00823 (Table 1), which is below the 0.05 threshold, suggesting that the CC distributions from the two methods originate from statistically different

populations. It is worth noting, however, that if a more stringent significance level of 0.001 (confidence level = 99.9%) is considered, the p-value exceeds this threshold. Under this criterion, the two CC distributions would not be considered significantly different, suggesting general consistency between the two correction approaches.

For the winter season, very high CC values were detected for regions in north of Midwest, east of Northwest, south and east of Southwest and central Arizona in the CONUS using both correction methods (Figs. 6e and 6f). Cold waves in these regions tend to occur frequently and remain localized, rarely influencing distant areas, an observation consistent with the high CC values indicating strong local interconnectedness (Bosikun et al., 2025).

Identifying grid points with high MGD values is particularly important to detect potential teleconnections, as they indicate long-range spatial dependencies in the occurrence of ETEs. The uncorrected and corrected MGD values for both the summer (JJA) and winter (DJF) seasons are presented in Fig. 7. The uncorrected results reveal MGD ranges from nearly 0 to 1350 km during summers (Fig. 7a) and from around 0 to 1500 km during winters (Fig. 7d). For the summer season, both correction methods yield similar spatial patterns, with notably high MGD values in regions, such as southern California, southeast Florida, and New England (Figs. 7b and 7c). These elevated MGD values suggest the presence of large-scale synchronizations and potential teleconnections in these areas.

Results of the paired t-test revealed a statistically significant difference between the mean MGD values computed using the subtraction and division correction methods, with a p-value much less than 0.05 (Table 1). Furthermore, the K-S test indicated that the MGD distributions obtained from the two methods are significantly different, confirming that they originate from distinct statistical populations (p-value << 0.05; Table 1).

The MGD results for the winter season are presented in Figs. 7d–7f. As shown in Figs. 7e and 7f, the surrogate-based subtraction and division methods produced visually similar corrections of the MGD. Via both approaches, we detected high MGD values, corresponding to teleconnections, in Texas, New England, and west of Northwest in the CONUS during winters (Figs. 7e and 7f). Cold air masses entering the CONUS through the Pacific Northwest during the winter season are often propagated southward and to the eastern parts (Colle & Mass, 1995). This indicates a long-range propagation of such air masses. The paired t-test yielded a p-value of 0.0834 (Table 1), indicating that the mean MGD values derived from the two correction methods are not significantly different at the 0.05 significance level. However, the K-S test produced a p-value much smaller than 0.05, suggesting that the MGD distributions from the two methods are statistically different significantly.

Consistent with the findings of Gupta et al. (2021), we observed notable similarities between the spatial distributions of DC and MGD across the CONUS during both summer and winter seasons (Figs. 5 and 7). Similar spatial correspondence between DC and MGD was also evident in the EPE networks (Figs. 1 and 3), suggesting a link between local connectivity and the spatial scale of synchronization in extreme events.

The MGD results presented in Fig. 7 revealed both regional connections and long-range teleconnections within the ETE networks. Notably, nodes with higher MGD values, indicating long-range synchronization, were more spatially widespread during winters (DJF) compared to summers (JJA). This pattern likely reflects the influence of larger-scale atmospheric systems, such as synoptic-scale disturbances and jet stream variability, which are more prevalent and spatially extensive in winters and contribute to the synchronized occurrence of ETEs across distant regions.

We present the spatial distribution of uncorrected and corrected BC values within the CONUS in Fig. 8 for the summer and winter seasons. The spatial distribution of uncorrected BC during the summer season (Fig. 8a) closely resembles that reported by Peron et al. (2014) who analyzed mean monthly temperature data across North America between 0° and 50°N (see their Fig. 3b). We should note that their network construction did not account for spatial boundary effects.

Similar to the BC results of the EPE network (Fig. 4), we found visually different results for the subtraction (Fig. 8b) and division (Fig. 8c) methods for the summer season. Results of statistical analyses also confirmed significant differences between the two correction methods. For both paired t-test and K-S test, we found p-values $\ll 0.05$ (Table 1).

Using the surrogate-based subtraction method, we generally found very low corrected BCs ($< 0.15$) everywhere in the CONUS (Fig. 8b). Exceptions are some regions in New York, Michigan, Arkansas, northeast Texas, north central Arizona, Nevada, Montana and southern California with high or relatively high BC values. Mondal & Mishra (2021) also identified southern California as a key origin point in the westward propagation pathways of heatwaves across the western U.S. (see their Fig. 2d). Heatwaves in the west of the CONUS have been shown to originate in southwestern coastal California, primarily due to adiabatic compressional heating and the transport of unusually warm air, and subsequently extend inland, affecting regions, such as the Central Valley, southern Nevada, and western Arizona (Lee & Grotjahn, 2016).

Using the surrogate-based division method, we identified a broader spatial extent of high BC values across the CONUS compared to the subtraction method (Fig. 8c vs. Fig. 8b). Specifically, regions such as New England, Texas (excluding eastern Texas), Arizona, southern California, Montana, Michigan, Colorado, and Arkansas exhibited elevated BC values. These areas may serve as critical pathways for the spatial propagation of ETEs. In a recent study, Agel et al. (2021) analyzed maximum daily temperature records from 35 stations across the northeastern U.S., including Maine, Rhode Island, New Hampshire, Massachusetts, Connecticut, Vermont, and New York, over the period 1980-2018. They defined heatwaves as periods of at least three consecutive days with maximum daily temperatures exceeding the 95$^{th}$ percentile threshold at each station. To explore the atmospheric drivers of heatwaves in the Northeast, Agel et al. (2021) incorporated additional datasets, including 500-hPa geopotential height, 2-m temperature and humidity, 900-hPa wind, and soil moisture. Using a k-means clustering approach, they identified four distinct circulation patterns associated with summer heatwaves in the region.

The BC results for the winter season, including both uncorrected and corrected values, are illustrated in Figs. 8d–8f. The uncorrected BC map (Fig. 8d) shows high values primarily in the southern Midwest and northern portions of the Southwest, especially Colorado, within the CONUS. Conversely, coastal areas and regions near the northern boundary with Canada exhibit notably low BC values, most probably due to limited connectivity imposed by spatial boundaries. The corrected BC spatial distributions, obtained via the surrogate-based subtraction and division methods, are shown in Figs. 8e and 8f, respectively. Similar to the results of the summer season (Figs. 8b and 8c), clear differences emerged between the two correction approaches. While both methods highlight high BC values across Texas, Oklahoma, eastern New Mexico, and eastern Colorado, they diverge in other parts of the CONUS. Notably, the division method (Fig. 8f) revealed a prominent pathway of elevated BC values extending from the Northwest along the eastern flank of the Rocky Mountains and another from the Northeast along the western side of the Appalachian Mountains, both converging toward Texas. This pattern suggests long-range structural importance of nodes in these corridors during wintertime ETEs propagation.

The results of the paired t-test and K-S test further confirm statistically significant differences between the two correction methods. Specifically, both tests yielded p-values well below the significance threshold of 0.05, indicating strong evidence against the null hypothesis. The paired t-test showed that the mean BC value derived from the subtraction method is significantly different from that obtained using the division method. Similarly, the K-S test revealed that the BC distributions corresponding to the two correction methods originate from statistically distinct distributions.

## 7. Limitations

In this study, we corrected climate networks of the EPEs and ETEs for spatial boundary effects using two widely applied methods: the surrogate-based subtraction and division approaches. We statistically compared the performance of these methods using the paired t-tests and K-S tests. Our results generally revealed significant differences between the two correction methods across multiple network measures i.e., DC, CC, MGD, and BC, for both summer (JJA) and winter (DJF) seasons.

This study, however, did not examine climate networks for the spring and fall seasons. Based on the results shown here, one should expect to see significant differences in the performance of the correction methods across network measures in these seasons as well. Further investigation is required to confirm and explore these potential differences.

Although Rheinwalt et al. (2012) qualitatively evaluated the subtraction method using a random network embedded on a sphere (see their Fig. 1), they did not perform statistical analyses to assess whether corrected metrics significantly differed from those derived at the global scale. Moreover, their evaluation was limited to a single network measure i.e., closeness centrality. Further investigations are still required to more comprehensively assess the performance of these correction methods, particularly by comparing regionally corrected networks with networks constructed at the global scale. However, generating global-scale climate networks at a $0.5° \times 0.5°$ spatial resolution would be computationally intensive due to the substantial number of nodes involved.

Overall, we found that the subtraction and division methods yielded statistically different outcomes, highlighting the importance of method selection when correcting for spatial boundary effects. Finally, we should note that our analysis was conducted using daily temperature and precipitation data and four widely used network measures; applying different temporal resolutions (e.g., hourly or monthly) and other network metrics (e.g., long-range directedness and Eigenvector centrality) may influence the outcomes and deserves future investigations.

## 8. Conclusions

In this study, we evaluated and compared two surrogate-based correction methods i.e., subtraction and division, for mitigating spatial boundary effects in climate networks of EPEs and ETEs over the CONUS during summers and winters. We used daily precipitation and temperature data from the CPC database at a spatial resolution of 0.5° × 0.5°. EPEs and hot ETEs were identified using the 95$^{th}$ percentile threshold, while cold ETEs were detected using the 5$^{th}$ percentile threshold. The ES method was used to quantify the similarity between extreme event series at each pair of nodes. Using the ES values, we constructed undirected and unweighted climate networks and computed four key network measures i.e., DC, CC, MGD, and BC. To correct for spatial boundary biases, we applied both the subtraction and division methods and statistically compared their outputs using the paired t-test and K-S test. Results showed that in the EPE networks, all corrected network measures significantly differed between the two methods (p-value << 0.05). In the ETE networks, statistically significant differences were also found for all network measures and seasons, except for the mean values of CC and MGD in winter, where p-values exceeded 0.05. Across the CONUS, the network hubs of EPEs were predominantly located in the northwestern United States during summer, shifting eastward in winter, consistent with seasonal variations in prevailing atmospheric circulation. In contrast, networks of ETEs exhibited stronger spatial coherence and more extensive teleconnections across both seasons, particularly in winter, when enhanced connectivity and longer synchronization distances reflected the influence of large-scale modes such as the Pacific–North American and North Atlantic Oscillation patterns. Further investigations are still required to evaluate how well these correction methods align with network structures constructed at the global scale.

**Acknowledgment**

BG is grateful to the University of Texas at Arlington for financial supports through faculty start-up fund and the STARs award. We acknowledge the use of ChatGPT 3.5 for language editing and proofreading of this manuscript. The AI was used to improve grammar, clarity, and readability. All intellectual contributions, data interpretation, and conclusions remain the sole responsibility of the authors.

**Data Availability**

The precipitation and temperature data used in this study were collected from the Climate Prediction Center (CPC) database, provided by NOAA/OAR/ESRL PSD (USA) and publicly available at https://psl.noaa.gov.

**Conflict of Interest**

The authors have no relevant financial or non-financial interests to disclose.

Rial, J. A., Pielke, R. A., Beniston, M., Claussen, M., Canadell, J., Cox, P., Held, H., De Noblet-Ducoudre, N., Prinn, R., Rynolds, J., & Salas, J. D. (2004). Nonlinearities, feedbacks and critical thresholds within the Earth's climate system. *Climatic Change*, *65*, 11–38.

Riedl, M., Marwan, N., & Kurths, J. (2015). Multiscale recurrence analysis of spatio-temporal data. *Chaos: An Interdisciplinary Journal of Nonlinear Science*, *25*, 123111.

Rosenberg, N. J. (1987). Climate of the Great Plains Region of the United States. *Great Plains Quarterly*, *7*(1), 22–32.

Singh, D., Swain, D. L., Mankin, J. S., Horton, D. E., Thomas, L. N., Rajaratnam, B., & Diffenbaugh, N. S. (2016). Recent amplification of the North American winter temperature dipole. *Journal of Geophysical Research: Atmospheres*, *121*(17), 9911–9928. https://doi.org/10.1002/2016JD025116

Svensson, C. (1999). Empirical orthogonal function analysis of daily rainfall in the upper reaches of the Huai River basin, China. *Theoretical and Applied Climatology*, *62*, 147–161.

Tadić, L., Bonacci, O., & Brleković, T. (2019). An example of principal component analysis application on climate change assessment. *Theoretical and Applied Climatology*, *138*, 1049–1062.

Trauth, M. H., Asrat, A., Duesing, W., Foerster, V., Kraemer, K. H., Marwan, N., Maslin, M. A., & Schaebitz, F. (2019). Classifying past climate change in the Chew Bahir basin, southern Ethiopia, using recurrence quantification analysis. *Climate Dynamics*, *53*, 2557–2572.
39

Table 1. p-values for statistical significance tests between distributions from the surrogate-based subtraction and division correction methods for various networks and network measures. KS represents the Kolmogorov-Smirnov test.

| Statistical test | Degree centrality (DC) | Clustering coefficient (CC) | Mean geographical distance (MGD) | Betweenness centrality (BC) |
|---|---|---|---|---|
| **EPE network-Summer (JJA)** | | | | |
| Paired t-test | 0.00 | 0.00 | 0.00 | 0.00 |
| KS test | 0.00 | 0.00 | $5.86 \times 10^{-83}$ | $6.16 \times 10^{-135}$ |
| **EPE network-Winter (DJF)** | | | | |
| Paired t-test | 0.00 | 0.00 | 0.00 | $5.17 \times 10^{-41}$ |
| KS test | 0.00 | 0.00 | $1.42 \times 10^{-60}$ | $3.15 \times 10^{-42}$ |
| **ETE network-Summer (JJA)** | | | | |
| Paired t-test | $3.71 \times 10^{-262}$ | $6.52 \times 10^{-19}$ | $1.93 \times 10^{-24}$ | 0.00 |
| KS test | 0.00 | 0.00 | 0.00 | 0.00 |
| **ETE network-Winter (DJF)** | | | | |
| Paired t-test | $1.19 \times 10^{-204}$ | $9.95 \times 10^{-1}$ | $8.34 \times 10^{-2}$ | $2.45 \times 10^{-290}$ |
| KS test | 0.00 | $8.23 \times 10^{-3}$ | 0.00 | 0.00 |

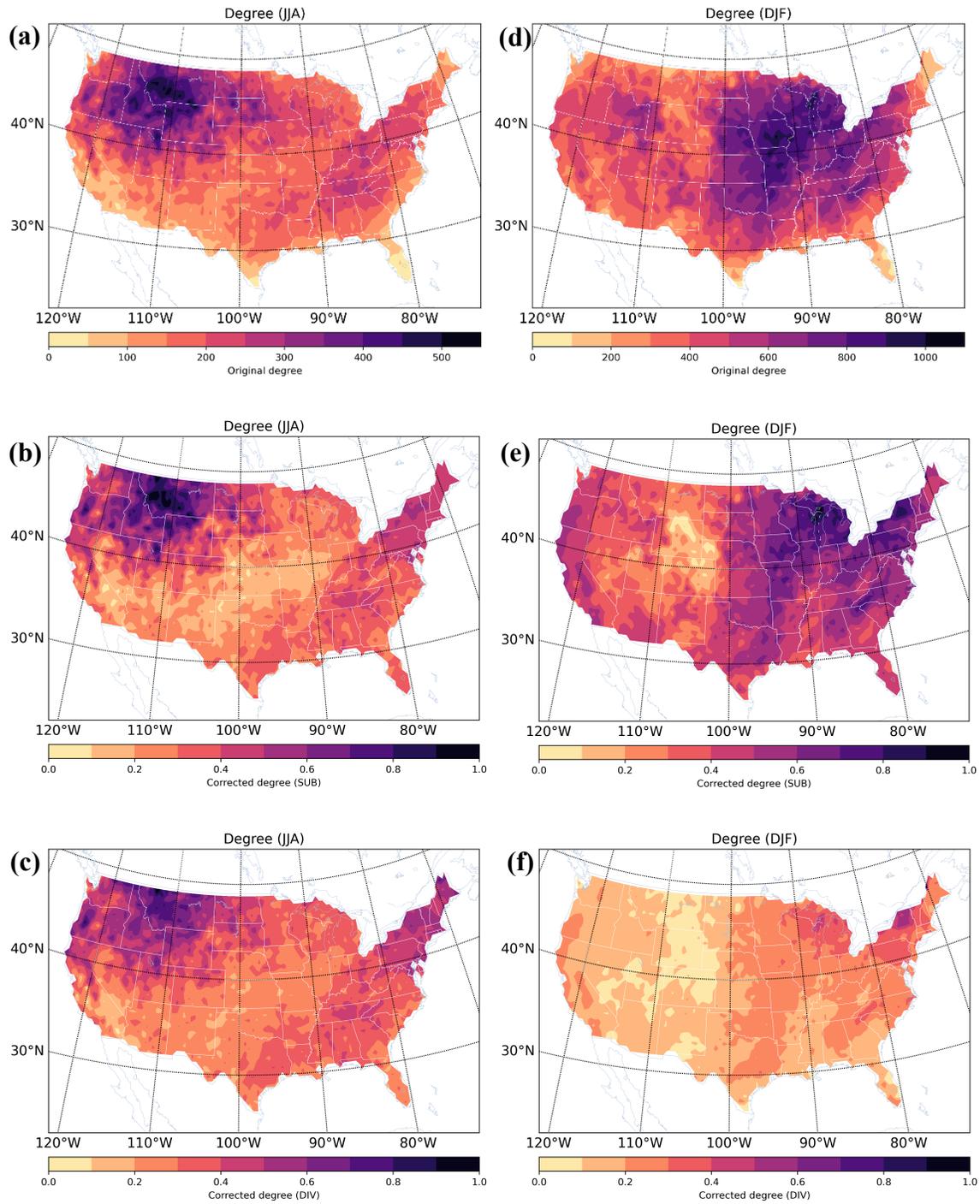

Fig. 1. Normalized degree centrality, DC, for the EPE network and (left) summer and (right) winter seasons. (Top) uncorrected, (middle) corrected via the surrogate-based subtraction method, and (bottom) corrected via the surrogate-based division method.

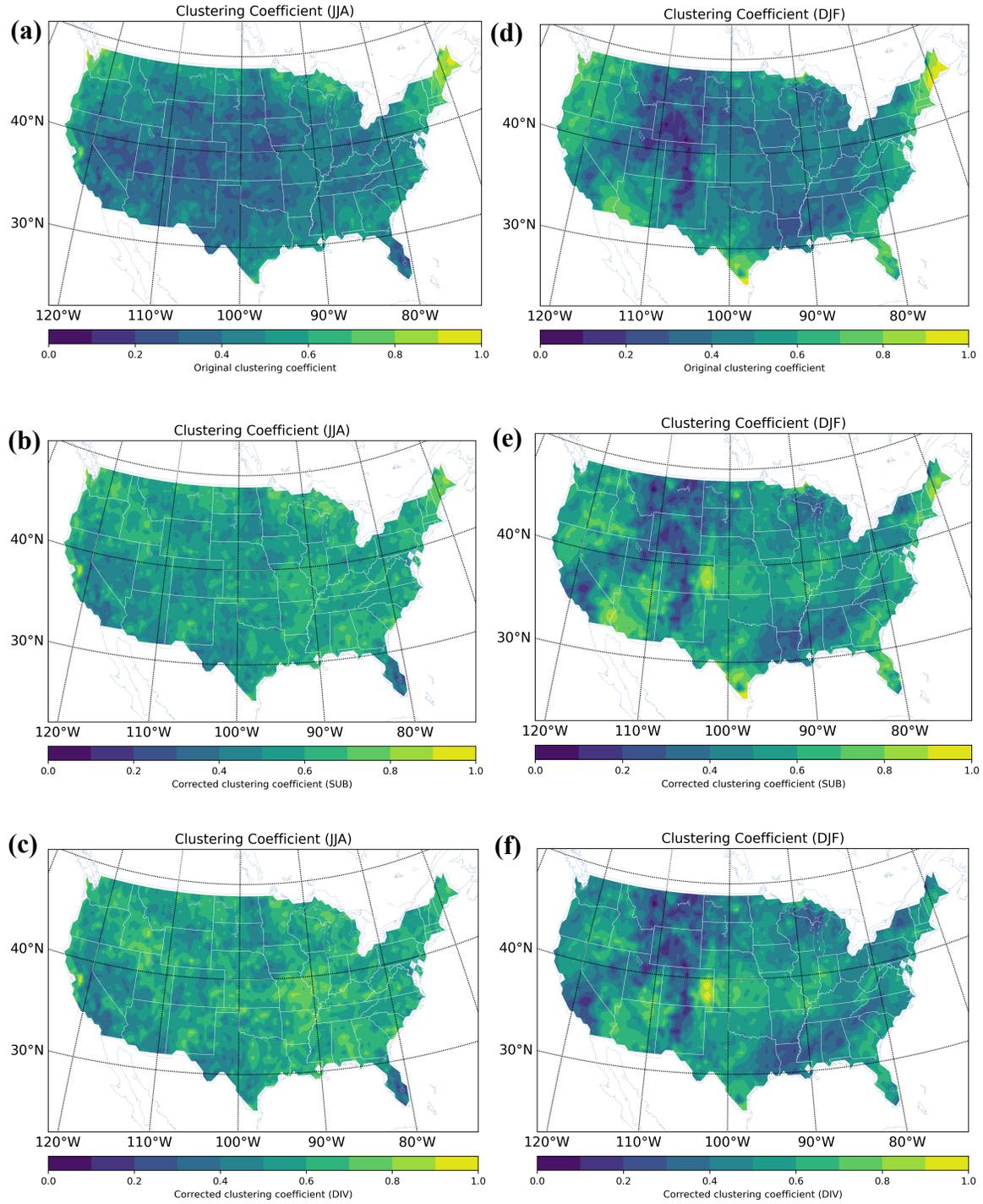

Fig. 2. Normalized clustering coefficient, CC, for the EPE network and (left) summer and (right) winter seasons. (Top) uncorrected, (middle) corrected via the surrogate-based subtraction method, and (bottom) corrected via the surrogate-based division method.

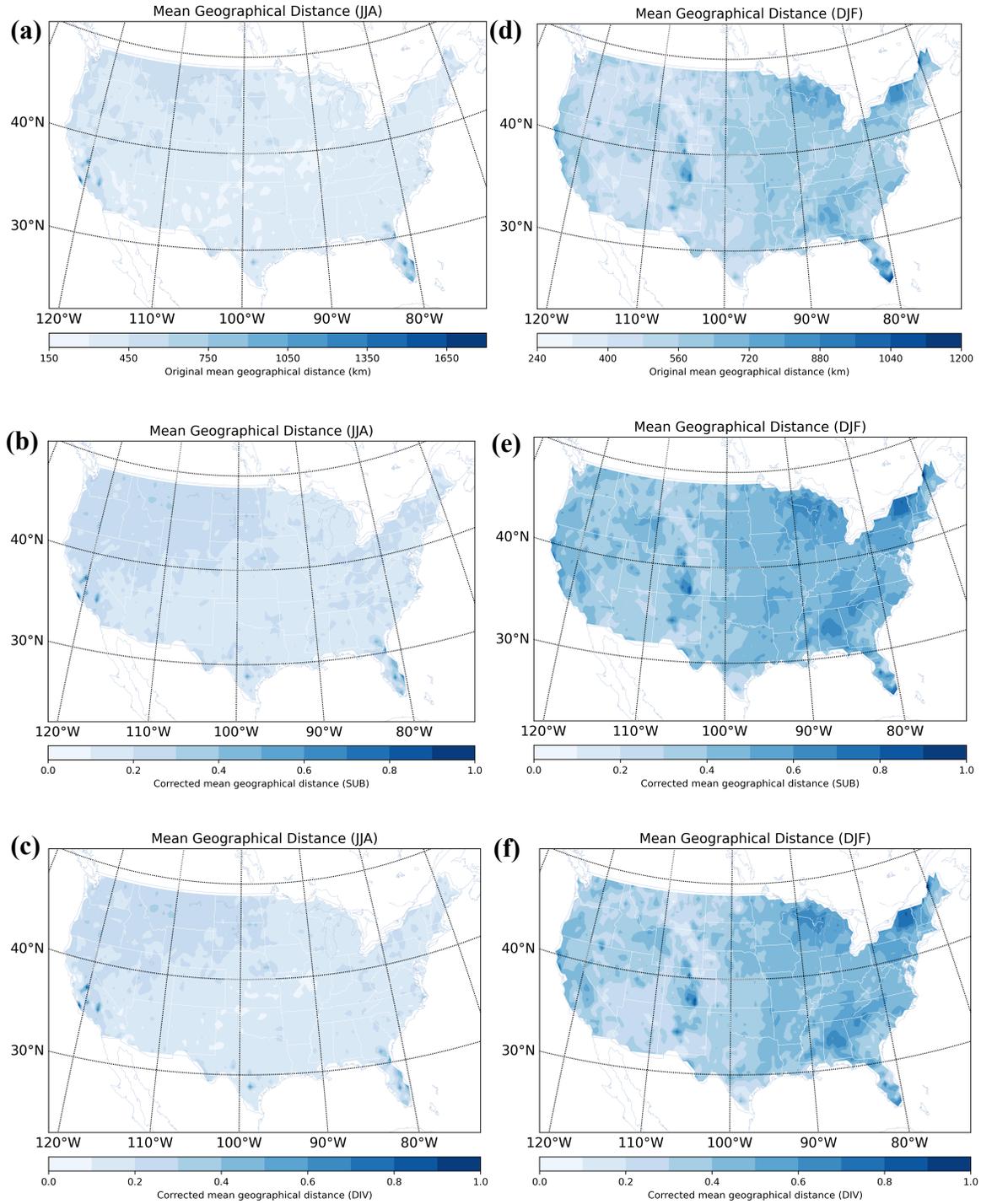

Fig. 3. Mean geographic distance, MGD, for the EPE network and (left) summer and (right) winter seasons. (Top) uncorrected, (middle) corrected via the surrogate-based subtraction method, and (bottom) corrected via the surrogate-based division method.

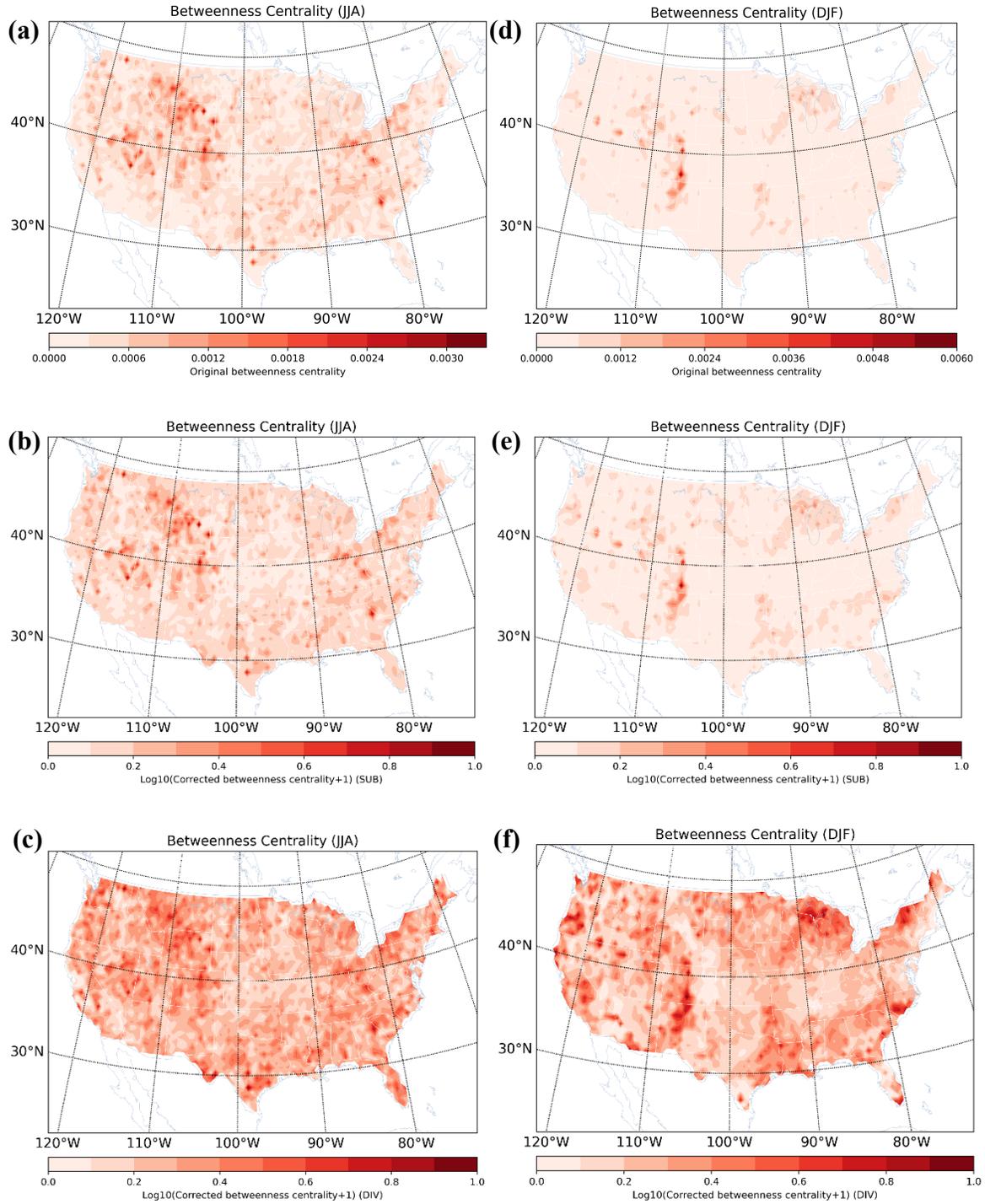

Fig. 4. Betweenness centrality, BC, for the EPE network and (left) summer and (right) winter seasons. (Top) uncorrected, (middle) corrected via the surrogate-based subtraction method, and (bottom) corrected via the surrogate-based division method.

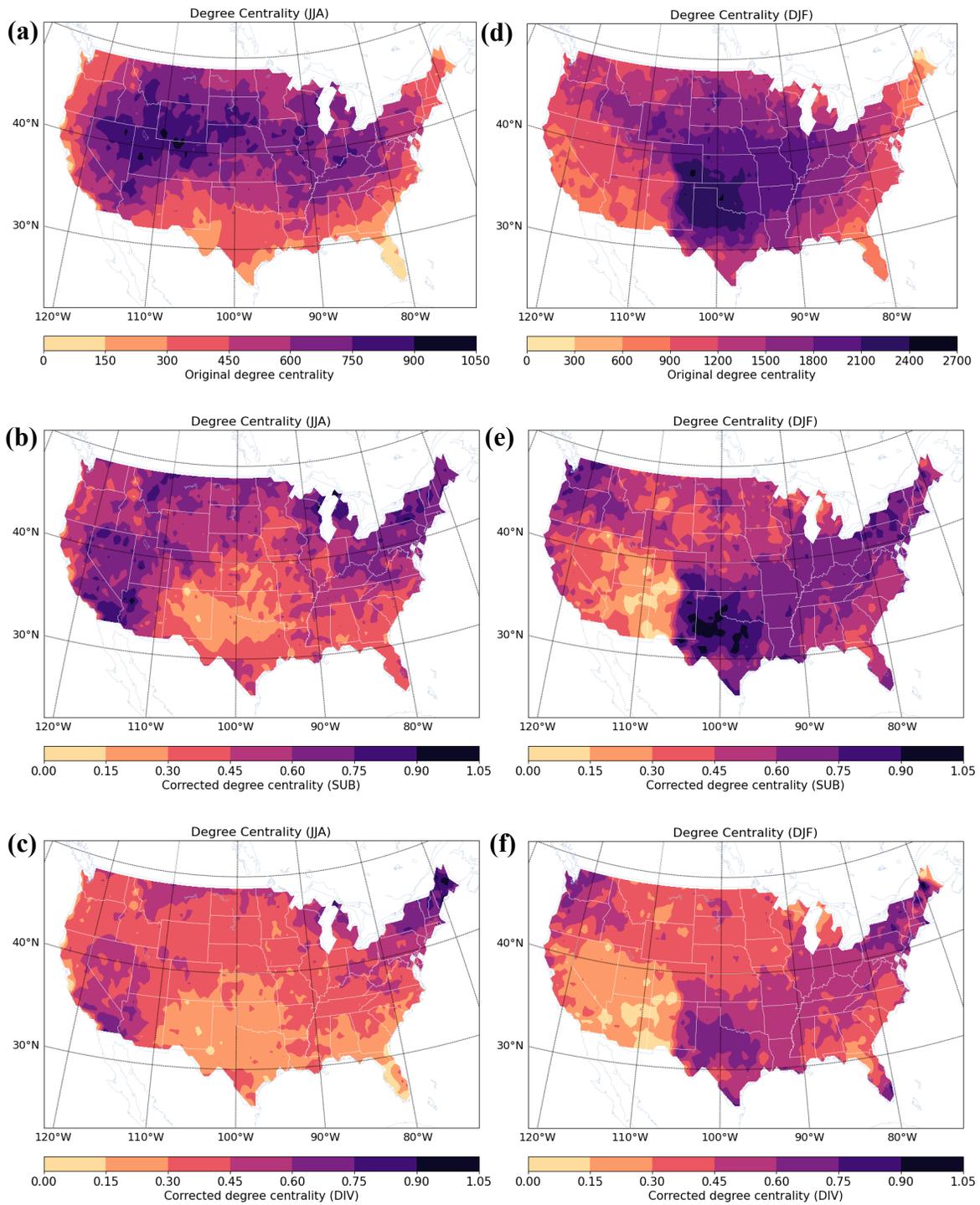

Fig. 5. Degree centrality, DC, for the ETE network and (left) summer and (right) winter seasons. (Top) uncorrected, (middle) corrected via the surrogate-based subtraction method, and (bottom) corrected via the surrogate-based division method.

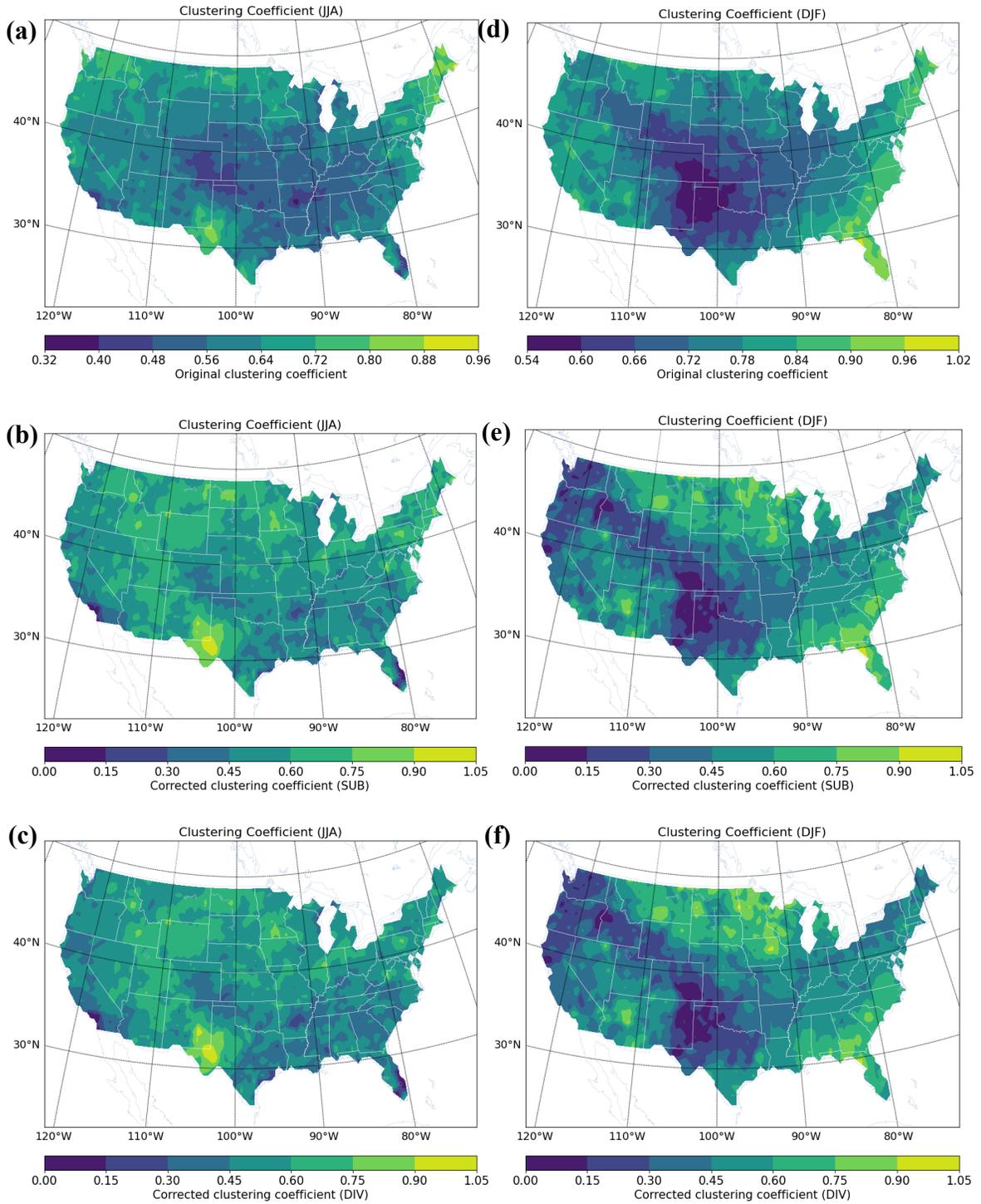

Fig. 6. Clustering coefficient, CC, for the ETE network and (left) summer and (right) winter seasons. (Top) uncorrected, (middle) corrected via the surrogate-based subtraction method, and (bottom) corrected via the surrogate-based division method.

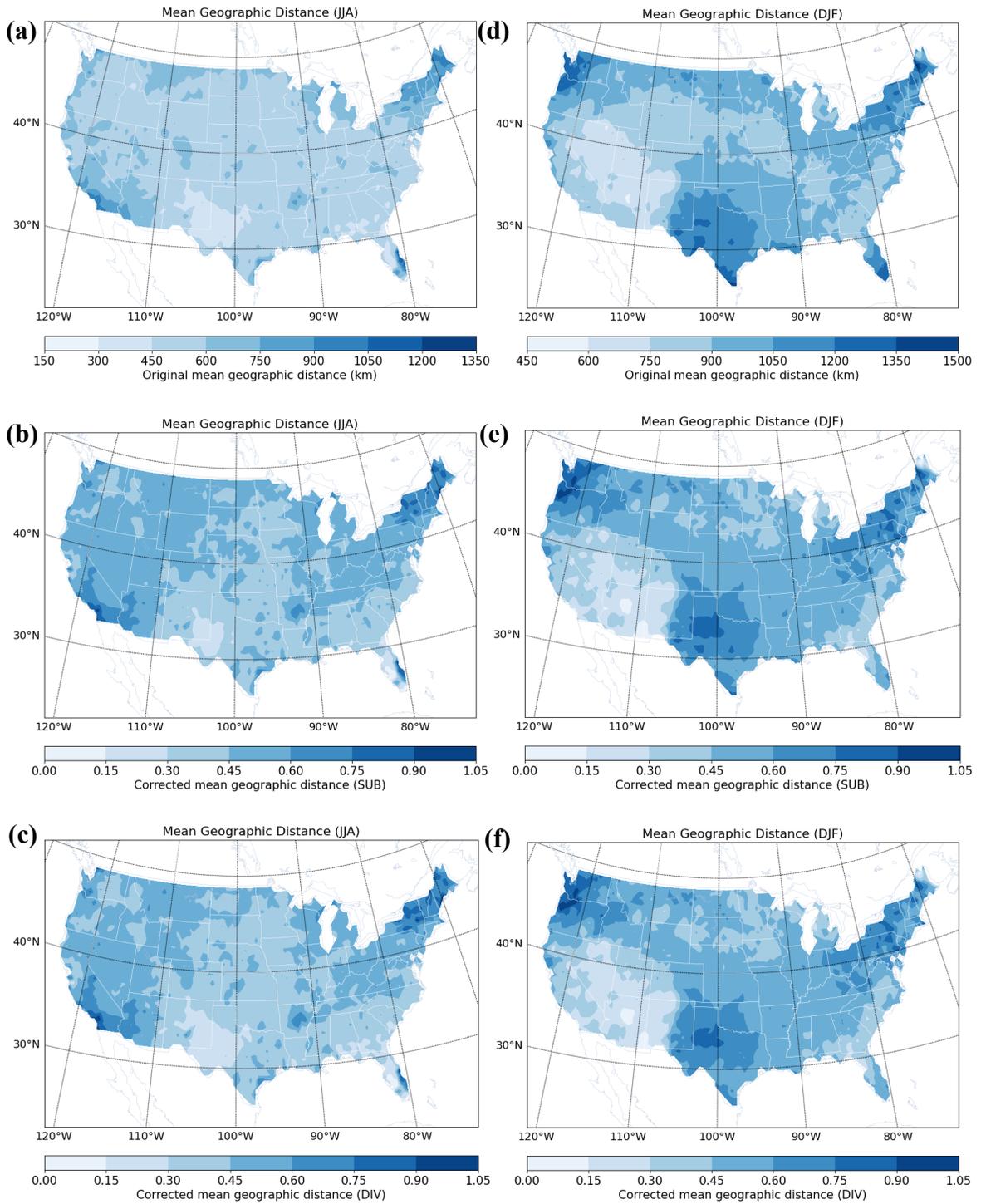

Fig. 7. Mean geographic distance, MGD, for the ETE network and (left) summer and (right) winter seasons. (Top) uncorrected, (middle) corrected via the surrogate-based subtraction method, and (bottom) corrected via the surrogate-based division method.

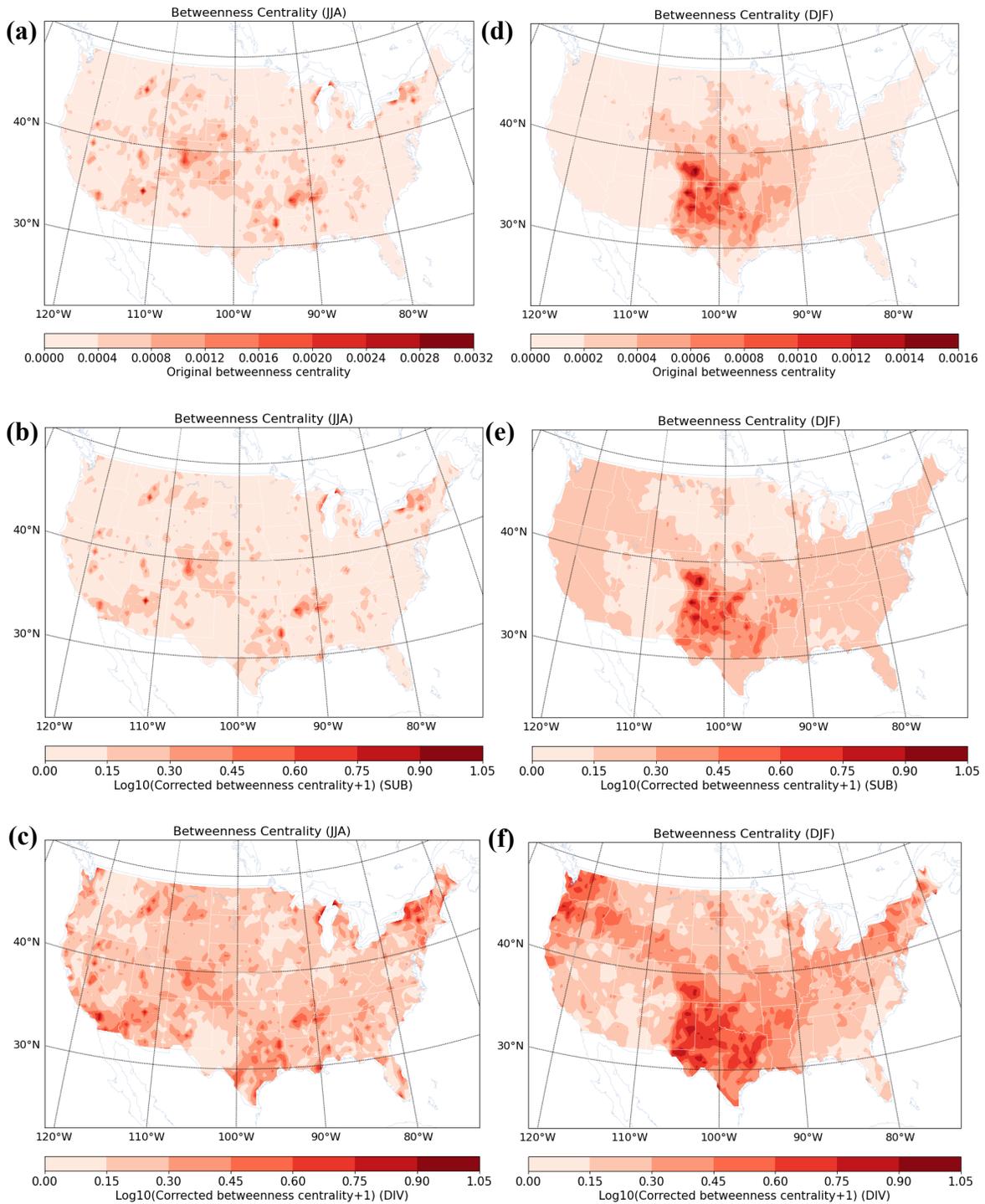

Fig. 8. Betweenness centrality, BC, for the ETE network and (left) summer and (right) winter seasons. (Top) uncorrected, (middle) corrected via the surrogate-based subtraction method, and (bottom) corrected via the surrogate-based division method.